\documentclass{emulateapj}

\def\Meszaros{M\'esz\'aros~}

\def\simg{\mathrel{\rlap{\raise 0.511ex \hbox{$>$}}{\lower 0.511ex \hbox{$\sim$}}}}
\def\siml{\mathrel{\rlap{\raise 0.511ex \hbox{$<$}}{\lower 0.511ex \hbox{$\sim$}}}}

\def\bLE{\beta_{LE}}     
\def\beg{\beta_{\eps\gamma}} 
\def\bog{\beta_{o\gamma}}

\def\tad{t_{ad}}         
\def\tsy{t_{sy}}         
\def\tsyi{t_{sy,i}}      
\def\tic{t_{ic}}         
\def\tici{t_{ic,i}}      
\def\tci{t_{rad}}        
\def\trad{t_{rad}}       

\def\tB{t_B}             
\def\ti{t_I}             
\def\to{t_o}             
\def\tang{t_{ang}}       

\def\gi{\gamma_i}        
\def\gc{\gamma_{cr}}     

\def\dtg{\delta t_\gamma}
\def\dto{\delta t_o}     
\def\dteps{\delta t_ \eps} 

\def\Ep{E_\gamma}        
\def\E5{E_{\gamma,5}}    
\def\Fp{F_p}             
\def\eps{\epsilon}       
\def\fe{f_\epsilon}      
\def\Fe{{\cal F}_\epsilon} 
\def\em{\varepsilon_m}   
\def\ep{\varepsilon_p}   

\def\te{t_{\gamma \epsilon}}   
\def\tp{\hat{t}_{\gamma \epsilon}}  
\def\tesy{\te^{(sy)}}    
\def\tead{\te^{(ad)}}    
\def\teic{\te^{(ic)}}    

\def\tgo{t_{\gamma o}}   
\def\tgosy{\tgo^{(sy)}}  
\def\tgoic{\tgo^{(ic)}}  
\def\tt{\tilde{t}}       
\def\tcr{t_{cr}}         

\def\tpk{t_p}            
\def\tpg{\tpk^{(\gamma)}} 
\def\tpe{\tpk^{(\eps)}}  
\def\Dte{\Delta t_\eps}  
\def\Dto{\Delta t_o}     

\def\epstoEp{\left(\h \displaystyle \frac{\eps}{\Ep} \h \right)}
\def\Eptoeps{\left(\h \displaystyle \frac{\Ep}{\eps} \h \right)}

\def\ds{\displaystyle}
\def\h{\hspace*{-1mm}} \def\hh{\hspace*{-2mm}} \def\hhh{\hspace*{-3mm}}

\def\ttimes{\h\times\h}

\begin{document}

\parskip 5pt
\topmargin 1cm

\title{Properties of the Prompt Optical Counterpart Arising from the Cooling of Electrons in Gamma-Ray Bursts}

\author{A.D. Panaitescu and W.T. Vestrand}

\affil{Intelligence and Space Research, MS D440, Los Alamos National Laboratory, Los Alamos, NM 87545, USA}

\begin{abstract}

 This work extends a contemporaneous effort (Panaitescu \& Vestrand 2022) to study the properties of the lower-energy 
 counterpart synchrotron emission produced by the cooling of relativistic Gamma-Ray Burst (GRB) electrons through 
 radiation (synchrotron and self-Compton) emission and adiabatic losses.
 We derive the major characteristics (pulse duration, lag-time after burst, brightness relative to the burst) 
 of the Prompt Optical Counterpart (POC) accompanying GRBs
 and whose short timescale variability indicates a common origin with the burst.

 Depending on the magnetic field life-time, duration of electron injection, and electron transit-time $\Dto$ 
 from hard X-ray (GRB) to optical emitting energies, a (true) POC may appear during the GRB pulse (of duration $\dtg$) 
 or after (delayed OC). The signature of counterparts arising from the cooling of GRB electrons is that true POC pulses 
 ($\Dto < \dtg$) last as long as the corresponding GRB pulse ($\dto \simeq \dtg$) while delayed OC pulses ($\Dto > \dtg$) 
 last as long as the transit-time ($\dto \simeq \Dto$). If OC variability can be measured, then another signature 
 for this OC mechanism is that the GRB variability is "passed" only to POCs but is lost for delayed OCs. 

  Within the GRB electron cooling model for counterparts, POCs should be on average dimmer than delayed ones 
 (which is found to be consistent with the data), and harder GRB low-energy slopes $\bLE$ should be associated more 
 often with the dimmer POCs.
 The latter sets an observational bias against detecting POCs from (the cooling of electrons in) GRBs with a hard 
 slope $\bLE$, making it more likely that the detected POCs of such bursts arise from another mechanism.

 The range of low-energy slopes $\bLE \in [-1/2,1/3]$ produced by electron cooling and the average burst brightness 
 of 1 mJy (with 1 dex dispersion) imply that POCs of hard GRBs can be dimmer than $R=20$ and difficult to detect 
 by robotic telescopes (unless there is another mechanism that overshines the emission from cooling electrons) 
 and that the POCs of soft GRBs can be brighter than $R=10$, i.e. as bright as the Optical Flashes (OFs) seen for 
 several bursts. 
 All GRBs with OFs identified in this work have a hard low-energy slope $\bLE$, thus these OFs were not produced by 
 the cooling of GRB electrons in a constant magnetic field.

 In many cases, the lag-time $\Dto$ between GRB and POC is about 100 times longer than the cooling timescale 
 of the GRB electrons, and that is the only new constraint on the GRB basic (non-temporal) model parameters that can be 
 extracted from observations of POCs, with the POC-to-GRB brightness ratio constraining two temporal model parameters.

\end{abstract}

\section{\bf Introduction}


 The {\sl prompt counterpart} is defined here as emission at an energy below the hard X-ray/$\gamma$-ray of the main burst, 
and whose short variability timescale suggests {\sl a common origin}. Simultaneity with the burst is not a requirement, 
as some prompt counterparts can appear during the burst (and are thus truly prompt), while other can appear after the burst. 
The latter will be called {\sl delayed counterparts}, to avoid the oxymoron delayed prompt counterpart, and should not be 
confused with the afterglow, whose lack of short timescale variability indicates that it arises from a different mechanism. 
 Given that a slowly-varying afterglow emission may exist even during the burst (e.g. GRB 050820A, whose OC was decomposed
by Vestrand et al 2006 into a brighter afterglow-like emission and a dimmer component varying synchronously with the GRB), 
the identification of the prompt counterpart (as defined above) should be based on its short variability timescale and, 
when possible, on its correlation with GRB pulses. 

 A short-lived {\sl Prompt Optical Counterpart} {\bf (POC)} emission has been detected during or after the prompt phase 
in many GRBs (e.g. sample listed by Kopac et al 2013) and could arise from these mechanisms: \\
$i)$ the {\sl reverse-shock} propagating in the GRB ejecta (\Meszaros \& Rees 1997, Panaitescu \& Meszaros 1998) 
 could produce a {\sl delayed} bright POC occurring {\sl after} the GRB pulse if its synchrotron emission peaks around 
 the optical, as proposed for the {\sl Optical Flash} {\bf (OF)} of GRB 991023 by \Meszaros \& Rees (1999) and Sari \& Piran (1999). \\
$ii)$ optical emission arriving {\sl before/during} the GRB pulse could be synchrotron emission from relativistic electrons, 
  which is upscattered (synchrotron self-Compton) to produce the GRB emission (Papathanassiou \& \Meszaros 1996, 
  \Meszaros \& Rees 1997), as proposed for the POC/OF of GRB 990123 by Panaitescu \& Kumar (2007). \\
$iii)$ {\sl prompt/delayed} OCs could arise from the {\sl pairs} (produced by GeV photons emitted during the GRB phase)
 that are accelerated by the forward-shock, as proposed for the OF of GRB 130427A by Vurm et al (2014), or which form in 
 the shocked medium, as proposed for same OF by Panaitescu (2015).

\begin{table*}[t]
\caption{\small \vspace*{2mm} \hspace*{10mm} Glossary of more frequently used notations} 
\vspace*{5mm}
\centerline{
\begin{tabular}{ll|ll}
  \hline \hline
Energies \\
\hline
$\gi$     & typical energy of injected electrons     &  $\eps$ & observing photon energy         \\
 \hline
Spectral quantities \\
\hline
$\bLE$    & GRB low-energy slope (below $\Ep$)       & $\bog$  &  optical-to-gamma effective spectral slope \\
$\Ep$     & peak energy of GRB $\nu F_\nu$ spectrum  & $\Fp$   &  flux at $\Ep$ \\ 
 \hline
Electron timescales \\
\hline
$\tad$    & AD cooling timescale                         & $\trad$ &  radiative cooling timescale of $\gi$ electrons \\ 
$\tsy$    & SY cooling timescale                         & $\tic$  & iC cooling timescale \\
$\tsyi$   & SY cooling timescale for the $\gi$ electrons & $\tici$ & iC cooling timescale for $\gi$ electrons \\
$\te$     & transit-time from GRB $\Ep$ energy to $\eps$ & $\tgo$    & transit-time from GRB to optical (1 eV)   \\
 \hline
Other timescales \\
\hline
$\tB$     & magnetic field life-time  & $\ti$  & electron injection duration \\
$\dtg$    & duration of GRB pulse     & $\dto$ & duration of POC pulse \\
$\tpk$    & pulse peak epoch          & $\Dto$ &  GRB-to-optical time-delay \\
$\tang$   & angular spread in photon arrival-time \\
  \hline \hline
\end{tabular}
}
\label{Notations}
\end{table*}

 Here, we study {\sl only the synchrotron emission} from cooled GRB electrons as a mechanism for POCs, for electrons cooling 
through synchrotron ({\bf SY}), inverse-Compton ({\bf iC}) (as synchrotron self-Compton), or adiabatic ({\bf AD}) losses. 
A POC dimmer than the burst
results if the magnetic field lives shorter than the radiative cooling timescale $\tci$ of the typical GRB electron. 
If injection of the GRB electrons stops before $\tci$ but their cooling continues until they radiate SY emission 
in the optical, then the POC peak-flux will be the same as that of the burst (assuming a constant magnetic field). 
However, if the electron injection lasts longer than the electron transit-time from gamma to optical emission, then a bright 
POC can be produced.

 We are interested in identifying temporal properties of POC resulting from the cooling of GRB electrons that can
be used as identifiers of this mechanism for POCs, albeit there are few criteria to discriminate the POCs arising from
the other three mechanisms above, most noteworthy being that \\
 $i)$ POCs produced by the reverse-shock and pairs formed from the GeV prompt emission should and could, respectively,
  occur after the GRB and \\
 $ii)$ POCs produced in the synchrotron self-Compton model for GRBs could occur before the prompt burst emission.

 {\bf Table \ref{Notations}} lists the most often notations used here.

\vspace{3mm}
\section{\bf Prompt Counterpart Properties}

 The analytical formalism for calculating the counterpart pulse light-curve $\fe(t)$ resulting from the cooling of GRB 
electrons is presented in a companion paper (Panaitescu \& Vestrand 2022 - PV22). Those light-curves allow the calculation 
of the counterpart peak flux and its brightness relative to that of the GRB, quantified by the effective GRB-to-counterpart
spectral slope $\beg$.
 
 The counterpart peak epoch $\tpk$ depends on \\
$i)$ the time $\dteps$ that it takes electrons "to cool through" the observing energy, which is often 
equal to the time $\te$ that it takes GRB electrons to migrate from $\gamma$-ray emitting energies to the energy $\eps$ at 
which the counterpart is observed, \\ 
$ii)$ the lifetime $\tB$ of the magnetic field and the duration $\ti$ of electron injection, \\
and, if the emitting surface is of uniform brightness, \\
$iii)$ the spread in photon arrival-time $\tang = R /(2c\Gamma)$ (in the source frame, with $R$ the source radius) across 
the ejecta surface of opening $\Gamma^{-1}$ (the inverse of the source Lorentz factor) from which the observer receives 
a relativistically enhanced emission.

 For pulses that are not too peaky or too stretched, the counterpart peak epoch $\tpk$ is also a good measure for the 
counterpart pulse duration $\dteps$. Another quantity of interest is the time-lag $\Dte$ between the GRB and the counterpart
peak epochs, which depends on the transit-time $\te$ and the two timescales $\tB$ and $\ti$ for the magnetic field 
lifetime and electron injection, but is independent on the angular spreading timescale $\tang$.

 For ease of access, we reiterate here some of the analytical results for the above counterpart properties of interest.
 
 If the pulse duration is set by radiative cooling (SY or iC through scatterings in the Thomson regime
having a cooling power $P_{ic} \sim \gamma^2$), then the $\gamma$-to-$\eps$ electron transit-time $\tesy$ is
\begin{equation}
  \tesy = \epstoEp^{-1/2} \tsyi 
\label{dtsy}
\end{equation}
Here, $\tsyi$ is the SY-cooling timescale of the typical GRB $\gi$ electrons that radiate SY emission at the peak energy $\Ep$ 
of the $\eps F_\eps$ GRB spectrum:
\begin{equation}
  \tsy (\gi) = \frac{\gi m_ec^2}{P_{sy}(\gi)} =  \frac{8.10^8 s}{B^2 \gi}
\label{gmsy}
\end{equation}

 For iC-cooling through scatterings at the Thomson--Klein-Nishina transition, characterized by a cooling power 
$P_{ic} \sim \gamma^{2/3}$, the electron transit-time is 
\begin{equation}
 \teic = \tici \left[1 - \epstoEp^{1/6} \right]
\label{dtic1}
\end{equation}
where $\tici$ is the iC-cooling timescale of the GRB $\gi$-electrons.

 For AD-dominated cooling and a power-law electron injection rate $R_i \sim t^y$
\begin{equation}
  \tead \simeq \epstoEp^{-3/4} \left\{ \begin{array}{ll} \ti & y < 1 \\ \to & y > 1  \end{array} \right.
\label{dtad} 
\end{equation}
where $\to$ is the epoch (ejecta age) when electron injection began.

\vspace{2mm}
\subsection{\bf Prompt Counterpart Timing}
\label{CPtiming}

 Equations (A8) and (A20) of PV22 for radiative electron cooling show that the pulse-peak epoch is at $\tpk = \te$ 
if the electron injection lasts shorter than the transit-time ($\ti < \te$) and for an exponent $n>1$ (as for SY 
and iC cooling in the Thomson regime) of the electron cooling power $P(\gamma) \sim \gamma^n$;
at $\tpk \in [\te,\te + \ti]$ if $\te < \ti$ for $n>1$, and at $\tpk = \te + \ti$ for $n < 1$.
Thus, $\tpk = \te + \ti$ is a good approximation for any ordering of $\te$ and $\ti$, 
a result that can be extended to AD cooling with $\te \equiv \tp^{(ad)}$ for $y < 1$ and 
$\te \equiv \te^{(ad)}$ for $y > 1$ (but only if $\te^{(ad)} > \ti$ in the latter case).

 The above results are valid if the magnetic field lives $\tB > \tpk$, i.e. if the electrons that yield the pulse peak
cool to below the observering energy.
 Conversely, for a short-lived magnetic field with $\tB < \tpk$, the pulse-peak epoch is set by $\tB$. 

 Thus, in general, the pulse peak-time is
\begin{equation}
 \tpe = \min \{\ti + \te,\tB \} + \tang  
\label{tpe}
\end{equation}
which, according to Equations (\ref{dtsy}), and (\ref{dtad}), also provides a good estimate for the 
pulse duration $\dteps$ for all cooling processes except iC-cooling dominated by scatterings at the T-KN transition ($n=2/3$), 
for which Equation (\ref{dtic1}) applies. That the peak-time $\tpk$ and pulse duration $\dteps$ are comparable is equivalent 
to a pulse rise and fall that are not too fast or too slow relative to the peak-time $\tpk$. 

 From Equation (\ref{tpe}), the {\bf time-delay (lag)} 
\begin{equation}
 \Dte \h = \tpe \h - \tpg \h = \min \{\te \h + \ti,\tB \} - \min \{\tci \h + \ti,\tB \}
\label{dtlag}
\end{equation}
between the burst and the lower-energy pulse-peak epochs is independent of the angular time-spread $\tang$. 
Here, $\tci$ is the radiative cooling timescale of the typical GRB $\gi$ electron: $\tci^{-1} = \tsyi^{-1} + \tici^{-1}$.

 {\bf Table \ref{ALL}} summarizes the temporal features expected for the GRB and lower-energy $\eps$ pulses resulting 
from the AD cooling or the radiative cooling with $n > 1$ of GRB electrons, for various orderings of the relevant 
timescales $\tci,\ti,\tB,\te$ and if the angular time-spread $\tang$ does not set the pulse duration.

 For a short-lived magnetic field $\tB < \ti + \tci$, pulses peak at $\tB$ (cases 1-3), and the GRB-to-low-energy 
lag-time $\Dte = 0$. For an electron injection duration satisfying $\te < \ti < \tB$, pulses peak at $\ti$ (case 4), 
yielding $\Dte = \te < \ti = \dtg$. In all these cases, the counterpart is truly {\bf Prompt}, defined by its peak occuring 
during the GRB pulse ($\Dte < \dtg$). 

 {\bf Delayed} counterparts, defined by the pulse-peak appearing after 
the GRB pulse ($\Dte > \dtg$), occur when the electron injection lasts $\ti$ shorter than the transit-time $\te$ 
and when the magnetic field $B$ is sufficiently long-lived $\tB > \max (\ti,\tci)$. 
 If SY emission stops before the transit-time (i.e. $\tB < \te$, cases 5-6), then the peak time-delay $\Dte$ is the
life-time $\tB$ of the magnetic field. 
 If SY emission is produced until after the transit-time (i.e. $\tB > \te$, cases 7-8), then the peak time-delay $\Dte$ 
is approximately the transit-time $\te$ given in Equation (\ref{dtsy}) for SY-cooling, in Equations (\ref{dtad}) for
AD-cooling, and the $\teic$ of Equation (\ref{dtic1}) for iC-cooling. 

 If the angular time-spread $\tang$ sets the pulse duration, then the duration of the GRB pulse $\dtg$ will be larger 
than given in {\sl Table} \ref{ALL} by at most a factor two. However, the angular time-spread $\tang$ does not affect 
the pulse-peak lag $\Dte$ because the GRB and low-energy pulse-peak epochs are delayed by the same duration $\tang$, 
thus the spherical curvature of the uniformly-emitting surface can change some delayed counterparts into prompt ones.

 As shown in {\sl Table} \ref{ALL}, prompt counterparts ($\Dte < \dtg$) should satisfy $\Dte < \dteps \siml \dtg$, 
i.e. {\sl prompt counterpart and GRB pulse durations are comparable} and delayed counterparts ($\Dte > \dtg$) should 
satisfy $\dtg < \dteps$ and $\Dte \siml \dteps$, i.e. a {\sl delayed counterpart lasts longer than the GRB pulse}. 

 {\bf Table \ref{TKN}} summarizes the temporal features of counterparts expected when the cooling of GRB electrons 
is dominated by iC scatterings at the T-KN transition, for which the cooling index is $n=2/3$. This cooling process yields 
POCs pulses satisfying $\tpk + \dteps = \tici + \ti$ (Equation \ref{dtic1}), which does not imply $\tpk \simeq \dteps$, 
thus the counterpart pulse duration $\Dte$ may not be comparable to its pulse peak-time $\tpe$.

\begin{table*}
 \caption{\small \vspace*{2mm}
  {\sl GRB and low-energy pulse properties} for an electron-cooling dominated by {\bf radiative losses} 
   with {\bf n $>$ 1} or by {\bf adiabatic losses}: 
  $i)$ pulse-peak epochs $\tpg$ and $\tpe$ 
  $ii)$ GRB-to-low-energy peak lag $\Dte \equiv \tpe - \tpg$ (which is independent of $\tang$), 
  $iii)$ low-energy-to-GRB slope $\beg$ for radiative-dominated electron cooling (dimmest counterpart is for $\beg=1/3$, 
       brightest for $\beg=-(n-1)/2$; $\beg \sim 0$ means that any value between those extremes is possible),
   for various orderings of the relevant timescales: 
   cooling $\tci$ of the GRB typical electrons, duration $\ti$ of electron injection, life-time $\tB$ of magnetic field, 
   electron transit-time $\te (> \tci)$ from GRB emission to observing energy $\eps$. 
    Counterparts are "very prompt" if $\Dte \ll \dtg$, {\bf prompt} if $\Dte \leq \dtg$, {\bf delayed} if $\Dte > \dtg$. 
    Counterpart brightness is relative to that of the GRB, with an "average" counterpart brightness corresponding to
   $\beg = 0$, i.e. a counterpart that is as bright as the burst (e.g. magnitude $R=16$ for a typical GRB peak-flux of 1 mJy).
   There is no clear correlation between the counterpart type and the counterpart-to-GRB brightness ratio.
   {\sl Note}: for AD electron cooling, $\tB$ sets the duration of SY emission but has no effect on electron cooling. 
   Furthermore, because the AD-cooling timescale is the current time, $\tci$ becomes $\ti$, thus only cases shown in 
   bold-face apply to AD-cooling. 
   {\sl Note}: the angular time-spread $\tang$ associated with the emission from a spherically curved surface increases 
   all timescales (including $\tpg$ and $\tpe$) by $\sim 50$\%, thus the type of OC does not change when going from
   a bright-spot to a spherical emitting surface. However, given that only electron cooling and angular integration
   induce the observed GRB pulse-duration decrease with increasing energy, the GRB pulses whose duration $\dtg$ is 
   set by $\tB$ or $\ti$ should be inconsistent with that trend if the SY emission arose from a bright-spot.
 }
\vspace*{5mm}
\centerline{
\begin{tabular}{clcccccccccc}
   \hline \hline
  & Case & $\tpg$   & $\tpe$ &  $\Dte$ & Counterpart &  $\beg$  & Counterpart \\
  &      & ($=\dtg$)  & ($=\dteps$)  &     &  Type   &          & Brightness  \\
   \hline
{\bf 1}& $\tB<\tci$        & $\tB$ &  $\tB$     &  0      & very prompt &  1/3       & dimmest    \\
2& $\tci<\tB<\te,\ti$& $\tB$ &  $\tB$     &  0      & very prompt & $\sim0$    & any        \\
3& $\tci<\te<\tB<\ti$& $\tB$ &  $\tB$     &  0      & very prompt & $-(n-1)/2$ & brightest  \\
4& $\tci<\te<\ti<\tB$& $\ti$ & $\ti+\te$  & $\te$   & prompt      & $-(n-1)/2$ & brightest  \\
   \hline
{\bf 5}& $\ti<\tci<\tB<\te$& $\tci$ &  $\tB$     & $\siml\tB$ & delayed  & $>0$       & dim        \\
{\bf 6}& $\tci<\ti<\tB<\te$& $\ti$ &  $\tB$     & $\siml\tB$ & delayed  & $\sim0$    & any        \\
{\bf 7}& $\tci<\ti<\te<\tB$& $\ti$ &  $\te$     & $\siml\te$ & delayed  & $<0$       & bright     \\
{\bf 8}& $\ti<\tci<\te<\tB$& $\tci$ &  $\te$     & $\siml\te$ & delayed  & $0$        & average    \\
   \hline \hline 
\end{tabular}
}
\label{ALL}
\end{table*}

\begin{table*}
 \caption{\small  \vspace*{2mm}
  {\sl GRB and low-energy pulse properties} for an electron-cooling dominated by {\bf iC-scatterings at the T-KN transition} 
   for which {\bf n = 2/3}. Because the transit-time $\te$ is shorter than the iC-cooling timescale $\tci$ of 
   the typical GRB electron, the ordering of $\tci$, $\ti$, and $\tB$ is not relevant.
   There is no correlation between the counterpart type and the counterpart-to-GRB brightness ratio.
   The counterpart-to-GRB slope is likely to be $\beg \simg 0$, thus these counterparts arising from an electron cooling 
   of exponent $n < 1$ are expected to be dimmer, on average, than those for an electron-cooling exponent $n > 1$ 
  ({\sl Table} \ref{ALL}). }
\vspace*{5mm}
\centerline{
\begin{tabular}{clcccccccccc} 
   \hline \hline
   &  Case              & $\tpg$    & $\tpe$    & $\Dte$ &   CP Type   &  $\beg$ & CP Brightness \\
   \hline
1& $\tB<\ti,\te$        & $\tB$     &  $\tB$    &   0    & very prompt & $< 1/3$ & dimmer   \\
2& $\te<\tB<\ti$        & $\tB$     &  $\tB$    &   0    & very prompt & $1/6$   & dim      \\
3& $\te<\ti,\ti+\te<\tB$& $\ti$  &  $\ti+\te$   &  $\te$ &  prompt     & 0       & average \\
   \hline
4& $\ti<\tB<\te$        & $\ti$     &  $\tB$  & $\siml\tB$ &  delayed  & $<1/3$  & dimmer   \\
5& $\ti,\te<\tB<\ti+\te$& $\ti$     &  $\tB$  & $\siml\tB$ &  delayed  & $<1/6$  & dimmer/average   \\
6& $\ti<\te,\ti+\te<\tB$& $\ti$     &$\ti+\te$  & $\te$    &  delayed  & 0       & average \\
   \hline \hline
\end{tabular}
\label{TKN}
}
\end{table*}

\vspace{2mm}
\subsection{\bf Prompt Counterpart Brightness (relative to GRB)}

 The pulse light-curves derived so far can be used to calculate the (peak-)flux at the counterpart pulse-peak and
to convert the ratio of counterpart-to-GRB peak fluxes (at different peak-times, separated by $\Dto$) to an effective
counterpart/optical-to-GRB spectral slope 
\begin{equation}
 \beg = \frac{\log \ds \frac{f_{pk}(\eps)}{f_{pk}(\gamma)} }{ \log \ds \frac{\eps}{\Ep}} \;,\;
 \bog = -0.2 \; \log \ds \frac{f_{pk}(1\,eV)}{f_{pk}(100\,keV)} 
\end{equation}

 If the electron cooling is SY-dominated (or dominated by iC-scatterings in the Thomson regime and with an index $n=2$), 
then the counterpart pulse light-curves given in equations (20)-(22) of PV22 lead to the OC-to-GRB slope 
\begin{displaymath}
  ({\bf SY/iC-Thomson})
\end{displaymath}
\begin{equation}
  \bog = \left\{ \begin{array}{ll}  
    \ds \frac{1}{3} & (1) \\ 
    \ds \frac{1}{3} - \frac{1}{3} \log \frac{\tB}{\tsyi}  & (2) \\ 
    \ds -\frac{1}{2}  & (3)\, (4) \\
    \ds \frac{1}{3} - \frac{2}{15} \log \frac{\tB}{\tsyi}  & (5) \\ 
    \ds \frac{1}{3} - \frac{2}{15} \log \frac{\tB}{\tsyi} - \frac{1}{15} \log \frac{\ti}{\tsyi}  & (6) \\ 
    \ds  -\frac{1}{5} \log \frac{\ti}{\tsyi}  & (7) \\ 
    0  & (8)    \end{array} \right.
\label{bogsy}
\end{equation}
for the cases listed in {\sl Table} \ref{ALL}.

 This equation shows that both prompt and delayed OCs can be either dim and bright relative to the GRB, depending on 
the ratios $\tB/\tsyi$ and $\ti/\tsyi$, without any expected correlation between the POC type and the POC-to-GRB
brightness ratio. Furthermore, for POCs with a slope $\bog \sim 0$ (i.e. between the extreme values $1/3$ and
$-1/2$), a measured slope $\bog$ constrains the ratio $\tB/\tsyi$, but the resulting constraint is unclear for delayed OCs. 

 When the electron cooling is AD-dominated, equations (B6) and (B8) of PV22 yield
\begin{displaymath}
  ({\bf AD: R_i \sim t^{-y}, y < 5/9})
\end{displaymath}
\begin{equation}
  \bog = \left\{ \begin{array}{ll}  
    \ds \frac{1}{3} & (1) \\ 
    \ds \frac{1}{3} - \frac{4}{45} \log \frac{\tB}{\ti}  &  (5)(6) \\ 
    0  & (7)(8) \\ 
       \end{array} \right.
\end{equation}
\begin{displaymath}
  ({\bf AD: 5/9 < y < 1})
\end{displaymath}
\begin{equation}
  \bog = \left\{ \begin{array}{lll}  
   \hh  \ds \frac{1}{3} - \frac{y-5/9}{5} \log \frac{\tB}{\to} & (1) \\ 
   \hh  \ds \frac{1}{3} + \frac{1-y}{5} \log \frac{\ti}{\to} - \frac{4}{45} \log \frac{\tB}{\to}  & (5)(6) & \hh  (\tB < \te) \\ 
   \hh  \ds \frac{3}{4}(1-y) - \frac{1-y}{5} \log \frac{\tB}{\ti}  & (7)(8) &  \hh (\te\h < \h\tB \h < \h\tp) \\ 
    0  & (7)(8) \\ 
       \end{array} \right.
\end{equation}
\begin{displaymath}
  ({\bf AD: 1 < y})
\end{displaymath}
\begin{equation}
  \bog = \left\{ \begin{array}{ll}  
    \ds \frac{1}{3} - \frac{4}{45} \log \frac{\tB}{\to} & (1)(5)(6) \\ 
       0  & (7)(8) \\ 
       \end{array} \right.
\end{equation}
with the corresponding cases of {\sl Table} \ref{ALL} indicated.
Thus, if the electron injection rate does not decrease too fast ($y < 5/9$), 
a measured optical-to-GRB slope $\bog$ constrains the ratio $\tB/\ti$ if $\tB > \ti$.

 For an electron cooling dominated by iC-scatterings at the T-KN transition, when the cooling index is $n=2/3$, 
equations (A18) and (A19) of PV22 lead to
\begin{displaymath}
  ({\bf iC/T-KN})
\end{displaymath}
\begin{equation}
 \hh  \bog = \left\{ \begin{array}{ll}  
  \hh \ds \frac{1}{3} + \frac{1}{5} \log \left(1-\frac{\tB}{\tic} \right)  & \hh (1) \\ 
  \hh \ds \frac{1}{6}  & \hh (2) \\ 
  \hh \ds 0            & \hh (3) \\
  \hh \ds\frac{1}{3} + \frac{1}{5} \log \left[\left(1-\frac{\tB}{\tic} \right) \left(1-\frac{\tB-\ti}{\tic} \right) \right] & \hh (4)\\ 
  \hh \ds \frac{1}{6} + \frac{1}{5} \log \left(1-\frac{\tB-\ti}{\tic} \right)  & \hh (5) \\ 
  \hh 0  & \hh (6)   \end{array} \right.
\label{bogic}
\end{equation}
where $\tic$ is the iC-cooling timescale of the $\gi$-electron and with the corresponding cases of {\sl Table} \ref{TKN} identified.
These results are valid for $\ti,\tB < \tic$. 
Similar to $n > 1$ electron cooling, for POCs, a measured slope $\bog$ represents a constraint on the ratio $\tB/\tic$.

\vspace{2mm}
\subsection{\bf GRB-to-Prompt Counterpart Lag-Time}

 From Equation (\ref{dtlag}), for a magnetic field with a life-time longer than the transit-time $\tB > \te$, 
the pulse-peak lag-time $\Dte = \te - \tci \siml \te$ is close to the transit-time $\te$ from GRB to the observing energy, 
given in Equations (\ref{dtsy}) for SY cooling, (\ref{dtad}) for AD cooling, and (\ref{dtic1}) for iC cooling at the
T-KN transition.

 Corrections occur when the transit-time $\te$ is sufficiently long that the electron-cooling departs from the cooling
laws used above. As shown in {\sl Appendix C} of PV22, if the cooling of the typical GRB $\gi$-electron is initially SY-dominated 
($\tsyi < 1.5\,\to$), it becomes AD-dominated after a critical time $\tcr$, but the 
evolution of the electron energy changes from the SY solution to 1/3-SY solution after a time equal to the initial 
ejecta age $\to$. Thus, for $\tesy \gg \to$, the correct transit-time $\tesy$ can be up to three times shorter than given 
in Equation (\ref{dtsy}).
 
 If the cooling of the GRB $\gi$-electron is initially AD-dominated ($\tsyi < 1.5\,\to$), then cooling remains AD-dominated 
at all times, yet a change in the electron cooling occurs at the epoch $\tt$ of equation (C12) of PV22, when the electron cooling 
switches from the AD-solution to the 1/3-SY cooling, leading to a more substantial correction for the transit-time 
$\tead$ if $\tead > \tt$. This is of relevance for an electron injection rate $R_i \sim t^{-y}$ with $y > 1$, 
when the pulse peak-epoch is set by passage of the lowest-energy electrons.

 For an electron injection rate with $y < 1$, the pulse peak-epoch is set by passage of the highest-energy electrons, 
whose cooling begins at time $\ti$, thus the initial cooling regime is set by the parameter 
$2\tsyi/3\ti = \tsyi/\tad(\ti)$, and with a corresponding $\to \rightarrow \ti$ substitution in the 
definition of the switch-time $\tt$. The calculation of the correct transit-time is further complicated if the
critical energy $\gc(t) \simeq (2/3)\gi\tsyi/(t+\to)$), where the AD and SY-cooling timescales 
are equal, crosses the observing energy $\eps$ before the high-end energy $\ep$ of the SY spectrum from the cooling-tail. 
As shown in fig 3 of PV22, the cooling-tail peaks below $\gc$ (but that peak is very shallow and broad) and the pulse-peak 
epoch corresponds to the time when the SY characteristic energy for the $\gc$-electrons crosses the observing energy $\eps$:
\begin{equation}
 \te^{(cr)} = \frac{2}{3} \left( \frac{\Ep}{\eps} \right)^{1/2} \tsyi = \frac{2}{3} \, \tesy
\end{equation}

 Putting together all above corrections, the lag-time between the GRB peak and the pulse peak-time at energy $\eps < \Ep$ is 
\begin{displaymath}
({\bf AD + SY}:  R_i \sim t^{-y}) \quad Z \equiv \frac{\tsyi}{\tad(t=0)}
\end{displaymath}
\begin{equation}
 \Dte = \left\{ \begin{array}{lll}
\hh \ds  \tesy            &       & \ds  Z \ll \epstoEp^{1/2}                   \\
\hh \ds  \frac{1}{3}\tesy &       & \ds  \epstoEp^{1/2} \ll Z < 1               \\
\hh \ds  \frac{1}{3}\tesy & \hh (y>1) & \ds 1 < Z \ll \Eptoeps^{1/4}    \\
\hh \ds  \tead            & \hh (y>1) & \ds \Eptoeps^{1/4} \ll Z        \\
\hh \ds  \te^{(cr)}       & \hh (y<1) & \ds 1 < Z < \frac{\ti}{\to},\Eptoeps^{\h 1/2} \hhh \ll 1 + \frac{\ti}{\tsyi}  \\
\hh \ds  \tp^{(ad)}       & \hh (y<1) & \ds \frac{\ti}{\to} < Z ,\Eptoeps^{\h 1/4} \hhh \ll 1 + \frac{\tsyi}{3\ti} \\
\hh \ds  \frac{1}{3}\tesy & \hh (y<1) & \ds 1 \h < \h Z \h < \frac{\ti}{\to},
        \Eptoeps^{\h 1/2} \hhh \gg 3\left(1\h+\h\frac{\ti}{\tsyi}\right) \\
\hh \ds  \frac{1}{3}\tesy & \hh (y<1) & \ds \frac{\ti}{\to} \h < \h Z ,\Eptoeps^{\h 1/2} \hhh \gg 
       \h \left(\frac{\tsyi}{3\ti}\right)^{\h 2} \hh + \h \frac{\tsyi}{3\ti}\h + 6 
       \end{array} \right.
\label{Dte}
\end{equation}
assuming that electron injection lasts longer than the initial AD cooling timescale ($\ti > \to$).
The above branches lack continuity whenever the result shown is asymptotic.

 By identifying the GRB-to-1-keV transit-time with those cases above that set a lower-limit on the observing energy $\eps$, 
one obtains the GRB-to-X-ray pulse-peak lag
\begin{equation}
 \Delta t_x \simeq \left\{ \begin{array}{lll}
 \hh \tesy \h = 10\, \E5^{1/2} \tsyi     &            & \h Z \ll 0.07 \E5^{-1/2}  \\
 \hh \tead \h = 30\, \E5^{3/4} \to      & \hhh (y>1) & \h Z \gg 6 \E5^{1/4}  \\
 \hh \te^{(cr)} \h = 7\, \E5^{1/2} \tsyi & \hhh (y<1) & \h \ds 1 \h < \h Z \h < \h \frac{\ti}{\to}, \frac{\ti}{\tsyi} \h \gg \h 10 \E5^{1/2} \\
 \hh \tp^{(ad)} \h = 3\, \E5^{1/4} \ti  & \hhh (y<1) & \h \ds \frac{\ti}{\to} \h < \h Z,  \frac{\tsyi}{\ti} \h \gg 3 \E5^{1/4} 
      \end{array} \right.
\label{Dtx}
\end{equation}
and those cases above that set an upper limit on $\eps$ should be identified with the GRB-to-POC pulse-peak lag 
\begin{equation}
 \Dto = \ds \frac{1}{3} \tesy \simeq 100 \, \E5^{1/2} \tsyi \equiv \tgosy
\label{tgosy}
\end{equation}
Thus, for all cases, the pulse peak-lag is $1/3$ of the SY transit-time $\tesy$ and
\begin{equation}
   \ds 0.01 \E5^{-1/2} \ll \frac{\tsyi}{\to,\ti} \ll 50 \E5^{1/4} 
\label{limits}
\end{equation}
is satisfied, with the ratio above being close to $\tsyi/\tad$, where $\tad = \to$ (initial age) for $y > 1$ (fast decreasing
electron injection rate) and $\tad = \ti$ (electron injection duration) for $y < 1$. 

 If the SY-cooling timescale $\tsyi$ of the GRB electrons is smaller than the lower limit above, then electrons are still in 
the SY-cooling regime by the time when they radiate in the optical, thus the transit time $\tesy$ will be three times longer 
than in Equation (\ref{tgosy}). If $\tsyi$ is longer than the higher limit above, then electrons are still in the AD-cooling
regime when they reach optically-emitting energies. The large gap of five decades between optical and GRB energies 
implies that, for GRB SY-cooling times $\tsyi$ ranging over almost four orders of magnitude, the electron cooling should be in 
the 1/3-SY cooling regime when the electron has cooled enough to radiate in the optical. Owing to its proximity to GRB energies, 
the peak-lag $\Delta t_x$ to soft X-rays (1 keV) has a more complex dependence on the initial cooling timescales 
(Equation \ref{Dtx}).

 The above results did not include iC cooling. When electron-cooling is iC-dominated by scatterings in the Thomson regime, 
the electron cooling-tail has a falling spectrum $\fe \sim \eps^{-1/2}$ and, as discussed in {\sl Appendix A1} of PV22, 
the POC pulse peaks at epoch $\te$ when the lowest energy electrons radiate in the optical, 
thus the gamma-to-optical transit-time is:
\begin{equation}
 ({\bf iC-Thomson}) \; \tgoic \h = \h (10^5\E5)^{1/2} \tici \simeq \frac{\tgosy}{Y}
\label{tgoic1}
\end{equation}
with $Y > 1$ the Compton parameter of the $\gi$ electrons, and $\tgosy$ the gamma-to-optical transit-time for SY-dominated 
electron cooling (Equation \ref{tgosy}). 
 This result is accurate if $\tic$ and $Y$ are constant, which is true if the condition for the growth of a power-law 
cooling-tail is satisfied. Otherwise, the Compton parameter above is an average during the electron cooling. 

 When electron-cooling is iC-dominated by scatterings at the T-KN transition, the cooling-tail has a rising spectrum 
$\fe \sim \eps^{1/6}$ and {\sl Appendix A2} of PV22 shows that the POC pulse peaks when the high-energy end of the cooling-tail 
falls below the optical (due to electron cooling after the end of electron injection at $\ti$) at epoch
\begin{equation}
 ({\bf iC-TKN}) \quad \tgoic \siml \ti + \tici = \ti + \frac{\tgosy}{300\, \E5^{1/2} Y} 
\label{tgoic2}
\end{equation}

 Equations (\ref{tgosy}), (\ref{tgoic1}), and (\ref{tgoic2}) give the GRB-to-POC pulse-peak delay for either the
emission from a bright-spot or a uniformly bright, spherically-curved surface because, in the latter case, all observer-frame 
timescales are stretched by the same angular spread in photon-arrival time $\tang$, thus the difference between the pulse-peak
epochs at two different observing energies should be unaffected by the time-spread $\tang$. 

 Over the visible surface of angular opening $\Gamma^{-1}$, the photons emitted from the edge (at angle $\theta = \Gamma^{-1}$ 
relative to radial direction) 
arrive at observer later by a duration $\tang$ than the photons emitted directly toward the observer (at angle $\theta=0$)
and have an energy in observer-frame that is twice smaller. This softening of the received emission due to the curvature
of the emitting surface will delay pulse peaks at lower energies, but the delay should be much less than $\tang$ because
the reduction in the relativistic boost by a factor 2 across the $\Gamma^{-1}$ region is raised to the third power in 
the received spectral flux (or flux density), thus the pulse-peak at a lower energy will occur well before $\tang$ after 
the peak at a higher ernergy. 

 Using Equation (\ref{tgosy}), the GRB-to-POC peak delay should be unaffected by the photon-energy and arrival-time 
spreads over the $\Gamma^{-1}$ visible region if $\tgosy \simeq 100\, \tsyi \simg \tang/{\rm few}$, which leads to
$\tsyi > 10^{-3} \to$ after using Equation (\ref{limits}). Thus, the result of Equation (\ref{tgosy}) provides a robust 
estimate of the GRB-to-POC peak-lag $\Dto$ for a sufficiently long-lived magnetic field ($\tB > \Dto$) and if electron 
iC-cooling is not dominant. Otherwise, the peak-lag time is that given in Equations (\ref{tgoic1}) and (\ref{tgoic2}).

\vspace{3mm}
\section{\bf Observational Constraints on Model Parameters}

 The GRB model has five {\sl basic} parameters: GRB ejecta Lorentz factor $\Gamma$, typical electron energy $\gi m_ec^2$, 
magnetic field $B$, total number of injected electrons $N_e$ (in the bright-spot or in the visible $\Gamma^{-1}$ region),
and source radius $R$, and two {\sl temporal} parameters: the timescales of electron injection $\ti$ and magnetic field $\tB$. 

\vspace{2mm}
\subsection{\bf Basic Model Parameters}

 The five fundamental parameters ($\Gamma,B,N_e,\gi,R$) determine the following observables: \\
 (1) the {\sl peak-energy of the GRB spectrum} 
\begin{equation}
 \Ep = \frac{2 \ttimes 10^{-8}}{z+1} B \Gamma \gi^2 \, (eV) 
\label{Ep0}
\end{equation}
thus a measurement of $\Ep$ sets this constraint on the basic model parameters
\begin{equation}
 B \Gamma \gi^2 = 5 \ttimes 10^{12}(z+1) \E5
\label{Ep1}
\end{equation}

 (2) the {\sl GRB-pulse duration} given in Equation (\ref{tpe}), with the transit-time $\te$ being the radiative cooling 
   timescale $\tci$. After taking into account that \\
$i)$ the angular spread timescale $\tang$ is negligible in the case of a bright-spot (of angular opening $\delta \theta 
  \ll \Gamma^{-1}$ much less than that of the area of maximal relativistic Doppler boost), \\
$ii)$ for the emission from uniform-brightness surface, the angular spread is $\tang = R/(2c\Gamma) = t_{co}/2$ is 1/3 
  of the AD-cooling timescale, \\
$iii)$ the AD-cooling timescale is $\tad = 1.5\,t_{co}$, \\ 
with $t_{co}$ the comoving-frame epoch corresponding to the end of electron injection or the disappearance of the magnetic field, 
the duration of the GRB pulse can be written as
\begin{equation}
  \dtg \h = \h \left\{ \begin{array}{ll} 
  \hh \min \{\ti \h + \tci(B;N_e,R),\tB \}                   & \hhh (\delta \theta \ll \Gamma^{-1} \h + Rad) \\
  \hh \min \{\ti \h + \tci,\tB \} \h + \tang \h              & \hhh (\delta \theta = \Gamma^{-1} \h + \h Rad) \\
  \hh \min \{2.5\,\ti,1.5\,\tB \} \simeq 2\,(t_{co} \h -\to) & \hhh (\delta \theta \ll \Gamma^{-1} \h + \h AD) \\
  \hh \min \{3\,\ti,2\,\tB \} \simeq 3\,(t_{co} \h -\to)     & \hhh (\delta \theta = \Gamma^{-1} \h + \h AD) \end{array}    \right. 
\label{dtg0}
\end{equation}
for either Radiative (SY and iC) or Adiabatic-dominated electron cooling. 
 For SY-dominated cooling, the radiative cooling timescale is $\tci \equiv \tsyi$ (Equation \ref{gmsy}); for iC-dominated cooling 
$\tci \equiv \tici = \tsyi/Y$ depends also on the number of electrons $N_e$ and source radius $R$ because they determine the electron 
scattering optical depth $\tau_e \sim N_e/R^2$ and the Compton parameter $Y (\gi, N_e,R) \sim \gi^2 \tau_e \sim \gi^2 N_e/R^2$. 

 From the first line of Equation (\ref{dtg0}), it can be inferred that
the GRB-pulse duration $\dtg$ is equal to the cooling timescale $\tci$ of the typical GRB electrons for the emission 
from a bright-spot, for a radiative-dominated electron-cooling, and for cases 5 and 8 of {\sl Table} \ref{ALL}.
In these cases, the constraint derived from the measured GRB pulse duration $\dtg^{(obs)} = (z+1)\dtg/\Gamma$ is
\begin{displaymath}
 ({\rm bright-spot \;\; \delta \theta \ll \Gamma^{-1} + Rad \;;\; case\; 5,8 - Table\, \ref{ALL}}) :
\end{displaymath}
\begin{equation}
 \ds B^2 \Gamma \gi (Y+1) = 8 \ttimes 10^8 \frac{z+1}{\dtg^{(obs)}} 
\label{dtg1}
\end{equation}
For a uniform-brightness surface and $\min(\ti,\tB)<\tang$, the GRB-pulse duration $\dtg$ (second line of Equation \ref{dtg0}) 
is set by the angular timescale $\tang$, thus a measured $\dtg$ implies 
\begin{displaymath}
 ({\rm unif-surface \;\; \delta \theta = \Gamma^{-1} + Rad \;\; or \;\; AD-cooling)}:
\end{displaymath}
\begin{equation}
 \frac{R}{\Gamma^2} = (1-6) \ttimes 10^{10} \frac{\dtg^{(obs)}}{z+1}
\label{dtg2}
\end{equation}
constraint which is also valid for the last two lines of Equation (\ref{dtg0}), corresponding to AD-dominated electron cooling,
if the comoving-frame source-age $t_{co} = R/c\Gamma$ is larger than the age $\to$ when electron injection began.

 (3) the {\sl average/peak SY flux} $\Fp$ at the GRB peak-energy $\Ep$
\begin{equation}
 \Fp = \frac{100\, Jy}{4\pi D_l^2/(z+1)} B \Gamma^3 N_e \min \left\{ 1, \frac{\tci}{\min(\ti,\tB)} \right\} 
\label{Fp0}
\end{equation}
where $D_l$ is the luminosity distance.
The last term above accounts for the cooling of electrons below the peak $\Ep$ (of the $\eps F_\eps$ power-per-decade): 
for $\tci < \min (\ti,\tB)$, only a fraction $\tci/\min(\ti,\tB) < 1$ of the total injected electrons $N_e$ radiate
at $\Ep$; and only a fraction $(Y+1)^{-1}$ is released as SY emission (that factor is ignored).
The last term above exists only for a radiative electron cooling because the AD-cooling timescale is the current 
time which implies
$\tci = \ti$ (with $\tB$ being irrelevant for electron cooling).

 The constraint provided by a measurement of the GRB peak-flux is
\begin{equation}
  B \Gamma^3 N_e = 10^{52}\frac{D^2_{l,28}}{z+1} \frac{\Fp}{mJy} \max \left\{ 1, \frac{\min (\ti,\tB)}{\tci(B,N_e)} \right\} 
\label{Fp1}
\end{equation}
where $D_{l,28}$ is the luminosity distance in units of $10^{28}$ cm. 
Note that the GRB peak-flux $\Fp$ does not depend on the temporal parameters $\ti,\tB$ {\sl only} if $\min (\ti,\tB) < \tci$,
which corresponds to cases 1, 5, and 8 of {\sl Table} \ref{ALL}, all cases of {\sl Table} \ref{TKN}, or if the electron cooling 
is AD-dominated.

 (4) the {\sl GRB-to-POC peak delay} $\Dto$ {\bf only} for cases 4, 7, and 8 of {\sl Table} \ref{ALL} and for cases 3 and 6 
of {\sl Table} \ref{TKN}, when $\Dto$ is equal to the gamma-to-optical transit-time $\tgo$, which is proportional to the
cooling-time $\tci(B;N_e,R)$ of the typical GRB electron. This is true whether the emission arises from a bright-spot 
or from a uniformly-bright surface because, for the latter, the angular time-spread $\tang$ delays equally both pulse-peak 
epochs. Furthermore, for SY-, AD-, and $n=2$ iC-dominated cooling, Equations (\ref{tgosy}) and (\ref{tgoic1}) can be combined 
to obtain the observer-frame GRB-to-POC lag-time $\Dto^{(obs)} = (z+1) \Dto/\Gamma$
\begin{displaymath}
 ({\rm case \; 4,7,8 - Table \; \ref{ALL}}) :
\end{displaymath}
\begin{equation}
  \Dto^{(obs)} \simeq 100\, \E5^{1/2} \frac{z+1}{\Gamma} \frac{\tsyi}{Y+1} 
\label{Dto0}
\end{equation}
leading to 
\begin{equation}
  B^2 \Gamma \gi (Y+1) \simeq 10^{11} (z+1) \frac{\E5^{1/2}}{\Dto^{(obs)}}
\label{Dto}
\end{equation}

 (5) the {\sl GRB-to-optical effective slope} $\bog$ between the peak fluxes of the GRB and POC pulses, in the cases indicated 
in Equations (\ref{bogsy}) -- (\ref{bogic}), by setting the cooling timescale $\tci$ of the typical GRB electron.

\vspace{2mm}
\subsection{\bf Radiative or Adiabatic Cooling ?}

 For the above cases (4, 7, and 8 of {\sl Table} \ref{ALL}, 3 and 6 of {\sl Table} \ref{TKN}) where the GRB-to-POC lag-time 
$\Dto$ provides a direct measurement (Equation \ref{Dto0}) of the radiative timescale $\trad = \tsyi/(Y+1)$ of the 
$\gi$-electrons, one can identify the radiative regime of those electrons using the AD-cooling timescale determined from 
the GRB duration (Equation \ref{dtg0}).

 But first, for a bright-spot emission and radiatively-cooling GRB electrons (first line of Equation \ref{dtg0}) and for 
case 8 of {\sl Table} \ref{ALL}, the GRB pulse duration is equal to the electron radiative-cooling timescale, $\dtg = \trad$, 
thus 
\begin{equation}
 \frac{\Dto^{(obs)}}{\dtg^{(obs)}} = 100\,\E5^{1/2} 
\end{equation}
Such long-delayed OCs are not found among the ten POCs of {\sl Table \ref{data}}. According to 
{\sl Table} \ref{ALL}, delayed OCs with $\Dto = \tgo$ are expected to have an average or above average brightness.
Then, the lack of long-delayed OCs could mean that radiatively-cooling electrons, a bright-spot emission, and 
$\ti < \trad < \tB$ occur in less than 10\% of POCs. However, if the cooling timescale $\trad$ of the GRB electrons 
is very short, then it would be difficult to identify the 10 ms GRB pulse corresponding 
to an POC peaking only 1 s after it.  

 For a uniform-brightness surface and radiatively-cooling GRB electrons (second line of Equation \ref{dtg0}), 
the GRB pulse duration is larger than the angular timescale ($\dtg \simg \tang$) and the condition for radiative 
electron-cooling ($\trad < \tad = 3\,\tang$) implies 
\begin{equation}
 \frac{\Dto^{(obs)}}{\dtg^{(obs)}} \siml 100\,\E5^{1/2} \frac{\trad}{\tang} < 300\,\E5^{1/2}  
\end{equation}
which is satisfied by all POCs of {\sl Table} \ref{data}. Thus, all POCs identified here {\sl could} be from radiatively 
cooling electrons (but that is not necessarily so).

 For an AD-dominated cooling of the $\gi$ electrons at the end of the GRB pulse, the GRB pulse duration (given in the 
last two lines of Equation \ref{dtg0}) is larger than the AD-cooling timescale: $\dtg = (2-3)\, t_{co} = (1.3-2)\,\tad$. 
As discussed in {\sl Appendix C} of PV22 and shown by some cases in
Equation (\ref{Dte}), depending on the $\tsyi/\tad(t=0) > 1$ ratio, the cooling-law of electrons may become a 1/3-SY
solution well before the AD-cooling electrons radiate in the optical, so that the gamma-to-optical transit-time $\tgo$ 
and the GRB-to-POC time-lag $\Dto$ are set by the SY-cooling timescale $\tsyi$ of the GRB $\gi$ electrons. 
In that case, the condition for AD-dominated electron cooling ($\trad(\gi) > \tad = 1.5\,t_{co}$) implies  
\begin{equation}
 \frac{\Dto^{(obs)}}{\dtg^{(obs)}} = 100\,\E5^{1/2} \frac{\trad}{(4-5)\,\tang} > (50-75)\,\E5^{1/2}  
\end{equation}
Such long-delayed OCs are {\sl not} found in {\sl Table} \ref{data}, which strengthens the previous suggestion that GRB 
electrons are cooling radiatively in at least 90\% of bursts, with the caveat that a long-delayed OC may be dimmer than 
indicated in {\sl Table} \ref{ALL} (cases 7 and 8 for AD-cooling), i.e. there could be an observational bias against
detecting them. 

 It is important to note that the above assessments are restricted to those cases where the radiative cooling timescale 
$\trad$ of the typical GRB electron can be measured from the GRB-to-POC time-lag $\Dto$ (Equation \ref{Dto0}) and where 
the AD timescale $\tad$ at the end of the GRB pulse (which is either the disappearance of the magnetic field at 
$\tB$ or the end of electron injection at $\ti$) can be constrained/determined from the duration $\dtg$ of the GRB pulse, 
using the second or the last two lines of Equation (\ref{dtg0}), respectively. 

\vspace{2mm}
\subsection{\bf Temporal Model Parameters}
\label{temppars}

 The two temporal parameters $\ti$ and $\tB$ for the duration of electron injection and magnetic field life determine: \\ 
$i)$ the GRB pulse duration $\dtg = \min (\ti, \tB)$, as shown for several cases in {\sl Tables} \ref{ALL} and \ref{TKN}), 
  {\sl only if} the SY emission arises from a bright-spot and that the electron-cooling is radiative-dominated (first line 
  of Equation \ref{dtg0}). In these cases, the $\dtg$ provides a direct measurement of either $\tB$ or $\ti$ (but the source
  radius $R$ remains unconstrained). \\
$ii)$ and the GRB peak (or average) flux $\Fp$ {\sl only if} $\tci < \min (\ti,\tB)$ (cases 2-7 in {\sl Table} \ref{ALL}). 
  In these cases, the $\Fp$ constrains the ratio $\tci/\min (\ti,\tB)$ (Equation \ref{Fp0}).  \\
$iii)$ the POC brightness relative to the GRB, quantified by the effective POC-to-GRB slope $\bog$ (Equations \ref{bogsy}--
  \ref{bogic}), for cases 2, 5, 6, 7 in {\sl Table} \ref{ALL} and cases 1, 4, 5 in {\sl Table} \ref{TKN}). In most of these 
  cases, the $\bog$ constrains either $\tci/\tB$ or $\tci/\ti)$.  

 Additionally, the magnetic field life-time $\tB$ sets the GRB-to-POC lag-time $\Dto$ for cases 5 and 6 in {\sl Table} \ref{ALL} 
and cases 4 and 5 in {\sl Table} \ref{TKN}. In these cases, the $\Dto$ provides a direct measurement of $\tB$, but there is
no overlap with a direct determination of $\tB$ from the GRB pulse duration $\dtg$ that could lead to a test of this model.
 
 Lastly, intermediate GRB low-energy slopes $\bLE \sim 0$ require field life-times $\tB$ that are just above the cooling timescale 
$\tci$ of the typical GRB electrons and should exhibit a good $\tB-\bLE$ correlation if $\tB \in (1,5)\trad$, for radiative electron 
cooling (equation 30 of PV22), and if $\tB \in (5\,\to,5\,\ti)$, for AD cooling (Equation 50 PV22).
These correspond to the very POC of case 2 in {\sl Table} \ref{ALL} and {\sl Table} \ref{TKN}, for which the GRB-to-POC 
time-lag $\Dto \sim 0$ does not constrain the electron cooling timescale $\tci$. 

 For these intermediate GRB low-energy slopes $\bLE \sim 0$, the power-law low-energy spectrum is not fully developed. 
Numerical calculations of that low-energy spectrum and fits to it with the Band function can quantify the $\tB/\tci-\bLE$
correlation and provide an observational constraint on the ratio $\tB/\tci$. 

 Measurements of the GRB low-energy slope could also be included in the determination of model parameters by assuming that 
the SY spectrum below the GRB peak-energy $\Ep$ is a perfect power-law of exponent $\bLE$, turning to a 1/3 slope below 
the minimal energy $\em$ reached by electron cooling after a duration $\tB$, and by using POC measurements during the GRB pulse 
(these POCs are very prompt). For example, for SY-dominated electron cooling, Equation (\ref{dtsy}) leads to
\begin{equation}
 \log \left( 1+\frac{\tB}{\tsyi} \right) = 2.5 \frac{1-3\,\bog}{1-3\,\bLE}
\end{equation}
which is consistent with the second line of Equation (\ref{bogsy}) for $\tB \in (1,300)\tsyi$.

 This result quantifies the $\bLE-\tB$ correlation of equation (20) in PV22 but includes POC measurements. Consequently, 
it does {\sl not} represent a substitute for the correlation that would be inferred from GRB observations alone (as described 
above) and is, instead, only a refinement of the second line of Equation (\ref{bogsy}), which assumed a GRB low-energy slope
$\bLE =-1/2$, and which is now set to the measured slope $\bLE$.

 We note that, in the framework of POCs arising from the cooling of GRB electrons, the duration $\dto$ of a POC  
follows from the duration $\dtg$ of the GRB pulse and that of a delayed OC follows from the GRB-to-POC peak-delay $\Dto$,
thus the measured POC pulse duration $\dto$ does not provide a constraint on the model parameters; instead, it can only
serve as a test of the POC origin in the cooling of GRB electrons. 

 Summarizing the above, we conclude that the GRB pulse duration $\dtg$ and GRB-to-optical time-lag $\Dto$ may constrain
directly the model temporal parameters $\tB$ and $\ti$, and the GRB low-energy slope $\bLE$ and GRB-to-optical slope
$\bog$ constrain the ratio of the temporal parameters to the electron cooling timescale $\tci$. For the cases listed
in {\sl Tables} \ref{ALL} and \ref{TKN}, the observables $\Dto$ and $\bog$ provide up to two constraints on the 
model temporal parameters, observable $\bLE$ provides another constraint only in case 2 (either Table) and a semi-constraint 
for all other cases, and observable $\dtg$ yields another constraint in most cases but only if the GRB emission arises 
from bright-spots and if electron cooling is radiative dominated. 

 We note that the equality of the two temporal parameters, $\tB = \ti$ (in a model where the production of magnetic fields
and the acceleration of relativistic particles are inter-dependent), selects cases shown in {\sl Tables} \ref{ALL} and 
\ref{TKN} for which $\tpg = \tpe$, i.e. only POCs. Thus, the existence of delayed OCs shows 
that the magnetic field life-time $\tB$ and the duration of electron injection $\ti$ are not always strictly equal.

\vspace{2mm}
\subsection{\bf Constraints on Basic Model Parameters}

 In the final tally, observations provide up to six constraints: $\Ep;\dtg,\Fp,\Dto,\bog;\bLE$ (first for basic parameters, 
next four for both basic and temporal parameters, last for temporal parameters) for seven model parameters: five basic 
($\Gamma,\gi,B,N_e,R$) and two temporal ($\tB,\ti$). 
 
 Even if one focused only on the conditions under which the temporal parameters $\tB$ and $\ti$ do {\sl not} determine 
the GRB pulse duration $\dtg$, the GRB peak-flux $F_p$, and the GRB-to-POC peak-delay $\Dto$ (i.e. only the emission 
from a bright-spot and case 8 of {\sl Table} \ref{ALL}, or the emission from AD-cooling electrons), one would still have 
only four constraints (Equations \ref{Ep1}, \ref{dtg2}, \ref{Fp1}, \ref{Dto}) for five model parameters. This system of 
four equations can be solved after choosing a free model parameter to parameterize the remaining four model parameters. 
Using the source Lorentz factor $\Gamma$ for that parameterization leads to
\begin{equation}
  \gi = 3 \ttimes 10^4\, \left(\frac{\Ep}{100\,keV} \right)^{1/2} \left[(z+1) \frac{\Dto^{(obs)}/10s}{\Gamma/100} \right]^{1/3} 
\end{equation}
\begin{equation}
  B = 50 \, \left[\frac{z+1}{(\Gamma/100)(\Dto^{(obs)}/10s)^2} \right]^{1/3} (G)
\end{equation}
for cases 4, 7, and 8 of {\sl Table} \ref{ALL} and for $Y < 1$, and
\begin{equation}
  R \siml (1-6) \ttimes 10^{15}\, \left(\frac{\Gamma}{100} \right)^2 \frac{\dtg^{(obs)}}{10(z+1)s} \; (cm)
\end{equation}
for the emission from a uniform surface ($\delta \theta = \Gamma^{-1}$) and radiatively-cooling electrons or for 
AD-dominated cooling.

 Constraints from the X-ray (1-10 keV) counterpart cannot break the degeneracy among the five model parameters because, 
for counterparts arising from the cooling of GRB electrons in a constant magnetic field, the prompt X-ray counterpart pulse 
duration, peak flux, and peak delay after to GRB should follow from the corresponding POC features. Conversely, the prompt 
X-ray and optical counterpart temporal and spectral properties not being consistent with each other would either constrain 
the evolution of the magnetic field or would indicate that the two emissions arise from different mechanisms.

 Thus, the full determination of the five model basic parameters requires the addition of another {\sl strong} observational 
constraint. Below, we discuss some weaker constraints that are either inequalities or rely on assuming some parameters
for the afterglow dynamics.

\vspace{1mm}
\subsubsection{Semi-Constraints}

{\sl From transparency to SY self-absorption.}
 The condition that the optical is above the SY self-absorption frequency (so that the optical continuum is not a hard 
$\beta_o = 2$) can be used to set a low-limit on the Lorentz factor: 
\begin{equation}
 \Gamma > (16-63) \left[ \frac{D_{l,28}^6}{(z+1)^4} \frac{ (F_p/mJy)^3}{\Dto^{(obs)}/10s} \right]^{1/8} 
\end{equation}
with the lowest value for SY-dominated electron cooling and emission from a uniform surface and the highest value for 
AD-dominated electron cooling.

{\sl From afterglow dynamics.}
 This constraint follows from the expectation that the GRB emission is produced before the interaction of the GRB ejecta
with the ambient medium leads to the deceleration of the post-GRB ejecta. 
Given that the dynamics of decelerating relativistic blast-waves, i.e. their radius $R_{bw}(t)$ and Lorentz factor 
$\Gamma_{bw}(t)$, are set by their kinetic energy and by the density of the circumburst medium, Kumar et al (2007) 
obtain upper-limits on the radius $R$ and lower-limits on the Lorentz factor $\Gamma$ of the GRB source using the
timing of the first afterglow measurements.


{\sl From escape of GeV photons.}
 The escape of 10-100 GeV photons requires a sub-unitary optical-thickness to pair-formation on the MeV burst photons 
(e.g. Abdo et al 2009), thus Fermi/LAT measurements of the GeV emission accompanying a GRB may set a stringent lower-limit 
$\Gamma_{min}$ on the source Lorentz factor, with the source whose radius $R$ determined from the GRB-pulse duration,
as in Equation (\ref{dtg2}).
 Evidently, this method provides an accurate lower-limit $\Gamma_{min}$ only if the GRB and GeV emissions arise from 
the same source, i.e. if the GeV emission occurs during the burst (temporal consistency) and if the GeV measurements
lie on the extrapolation of the burst MeV spectrum above the peak-energy $\Ep$ or have a spectral energy distribution
consistent with that of the burst (spectral consistency).

\vspace{3mm}
\section{\bf Discussion}

\vspace{2mm}
\subsection{\bf Temporal Correlations between GRB and POC}
\label{tempcorrel}

 By comparing the GRB-to-POC pulse-peak delay/lag $\Dto$ with the GRB pulse duration $\dtg$, one can identify 
two types of POCs. {\sl Prompt} POCs, defined by $\Dto < \dtg$ (i.e. POC pulse-peak occurs {\sl during} the GRB pulse), 
arise when the pulse-peak epochs (or pulse durations) given in Equation (\ref{tpe}) are both determined either 
by the duration of electron injection $\ti$ or by the life-time $\tB$ of the magnetic field. 
{\sl Delayed} POCs, defined by $\Dto > \dtg$ (i.e. POC pulse-peak occurs {\sl after} the GRB pulse) occur whenever
the two pulse durations are determined by different factors that introduce different timescales. 

 The GRB and POC temporal features of {\sl Tables} \ref{ALL} and \ref{TKN} refer to the case when the spread $\tang$ 
in the photon arrival-time 
over the curved emitting surface does not set the pulse duration. If the electron cooling is dominated by a radiative
process (i.e. $\tsyi,\tici < \tad = 3\,\tang$), that assumption is correct only if the emitting region is a bright 
spot of angular extent less than the $\Gamma^{-1}$ region (moving toward the observer) of maximal Doppler boost. 
The same assumption is correct if the electron cooling is dominated by AD losses because, in that case,
the comoving-frame angular timescale $\tang$ is less than the comoving-frame current age, which is comparable to 
the dominant timescale appearing in Equation (\ref{tpe}) for the pulse-peak epoch and pulse duration.

\begin{table*}[t]
 \caption{\small  \vspace*{3mm}
 {\sl Temporal and spectral properties of some GRBs with Prompt Optical Counterparts.}
 The burst pulse number and the GRB instrument are indicated (KW=Konus-Wind). 
 The $\tpg$ is the GRB peak epoch measured from trigger (not from the beginning of the pulse) and is of no use, 
  but is listed here for identifying the GRB pulse.
 GRB pulse duration $\dtg$ and optical pulse duration $\dto$ are the width at half-maximum read from the light-curves 
  found in References; $\Dto$ is the GRB-to-POC peak delay/lag-time; their uncertainties are 10-20\%. 
 The GRB low-energy slope $\bLE$ is taken from GCN circulars and has an uncertainty of at least 0.1.
 The effective slope $\bog$ between the optical and $\gamma$ pulse-peak fluxes (separated by $\Dto$) was calculated 
  from peak fluxes found in Reference, and has an uncertainties less than 0.1. 
 {\bf Prompt OCs} are defined by $\Dto < \dtg$ (optical peak occurring during the GRB pulse), 
 {\bf Delayed OCs} are defined by $\Dto > \dtg$ (optical peak occurring after the GRB pulse). 
 One expects that $\dto \simeq \dtg$ for POCs and $\dto \simeq \Dto$ for delayed OCs, 
 For either type, it is also expected that $\bLE \leq \bog$, with the equality taking place for $\bLE = \bog = 1/3$ or -1/2. 
 {\bf Optical Flashes} are counterparts with an optical emission brighter than the expectation for the cooled 
  GRB electrons (in a constant magnetic field) and satisfy $\bLE > \bog$.
  As shown in Figure 1, most OFs do not display the above temporal correlations expected for POCs (if $\Dto < \dtg$ then 
  $\dto \simeq \dtg$) or for Delayed POCs (if $\Dto > \dtg$ then $\dto \simeq \Dto$).
 Evidently, there is an observational bias in favor of detecting OFs using robotic telescopes than following 
 up dim POCs with $\bog = 1/3$, which implies that the measured slopes $\bog$ are sometimes softer than expected for the 
 cooling of GRB electrons. 
 }
\vspace*{5mm}
\centerline{
\begin{tabular}{ccccccccccc}
   \hline \hline
   GRB    &  Pulse  & $\tpg$ &                 &       &       &                          &        &  Reference   \\
   \hline 
\multicolumn{2}{c}{(Cooling) Prompt OC} & (s) & \multicolumn{3}{c}{$\dtg\simeq\dto>\Dto$} & \multicolumn{2}{c}{$\bLE\leq\bog$}   &  \\
   \hline 
  060526  & 2 (BAT) &  255  &             30   &    35 &   10  &                    -0.5  &  0.0   &  Thone 2010, Kopac 2013 \\
  061121  & 3 (KW)  &   72  &              8   &    10 &    5  &                    -0.3  &  0.1   &  Page 2007 \\
  110205  & 3 (BAT) &  210  &             20   &    15 &    5  &                    -0.5  &  0.2   &  Cucchiara 2011, Gendre 2012  \\
  120711  & 2 (IBIS)&   90  &             25   &    30 &   20  &                    -0.4  & -0.5   &  Martin-Carrillo 2014 \\
   \hline 
  130925  & 2 (KW)  & 2650  &            280   &   300 &  290  &                    -0.4  &  0.0   &  Greiner 2014 \\
   \hline 
\multicolumn{2}{c}{(Cooling) Delayed OC}&    & \multicolumn{3}{c}{$\dtg<\dto\simeq\Dto$} & \multicolumn{2}{c}{$\bLE\leq\bog$}    &   \\
   \hline 
  041219  & 3 (BAT) &  430  &             15   &    30 &   35  &                    -0.7  &  -0.4  &  Vestrand 2005, Blake 2005 \\
  060904B & 1 (BAT) &    2  &              4   &    70 &   50  &                    -0.5  &   0.1  &  Klotz 2008, Kopac 2013 \\
  091024  & 1 (KW)  &    5  &             15   &   450 &  420  &                    -0.5  &  -0.5  &  Gruber 2011 \\
  091024  & 5 (KW)  &  930  &            130   &  1400 & 1800  &                    -0.4  &  -0.4  &  Virgili 2013 \\
  111209  & 3 (KW)  & 2000  &            350   &   500 &  450  &                    -0.3  &   0.0  &  Stratta 2013 \\
  \hline
\multicolumn{2}{c}{Optical Flashes}&  & $\dtg$ & $\dto$  & $\Dto$   & \multicolumn{2}{c}{$\bLE > \bog$}     &          \\
  \hline
  990123  & 1 (BATSE) &  25 &             10   &   40? &   22? &                    0.4   &  -0.6  &  Akerlof 1999 \\
  990123  & 2 (BATSE) &  37 &             10   &   40? &   10? &                    0.0   &  -0.6  &  Corsi 2005 \\
  080319B & 1 (BAT)   &  18 &             22   &   25  &    2  &                    0.15  &  -0.7  &  Racusin 2008, Wozniak 2009 \\
  121217  & 2 (GBM)   & 730 &              8   &   45  & 10-20 &                    0.4   &   0.1  &  Elliott 2014 \\
  130427A & 2 (GBM)   &  8  &              6   & $<12$ &$\sim5$&                    0.0   &  -0.5  &  Ackermann 2014, Vestrand 2014 \\
  160625B & 1 (KW,GBM)& 195 &             20   &   20  & 10-20 &                    0.2   &  -0.4  &  Karpov 2017, Ravasio 2018 \\
   \hline \hline
\end{tabular}
}
\label{data}
\end{table*}

 For a radiative electron cooling ($\tci < \tad$) and an uniformly-emitting surface, the peak epochs and 
pulse durations are increased by the angular time-spread $\tang$, but the pulse-peak lag-time $\Dto$ 
(Equation \ref{dtlag}) remains unchanged. Consequently, including the angular time-spread $\tang$
in the peak-epoch and in the pulse duration will change some $\tang$-free delayed OCs with $\Dto > \dtg$ into 
$\tang$-included POCs with $\Dto < \dtg$, as the GRB pulse duration is increased. 

 With that limitation for the identification of POC types, one can search for correlations between the temporal/spectral 
properties of GRBs/POCs for these two types of POCs. 

 {\bf Figure 1}, showing the OCs listed in {\bf Table \ref{data}}, illustrates the temporal {\sl correlation between 
POC type and optical-to-GRB pulse duration ratio $\dto/\dtg$} discussed in \S\ref{CPtiming}, which can be summarized as
\begin{equation}
 \left\{ \begin{array}{ll}  {\rm prompt\, OC:} & \Dto < \dto \simeq \dtg \\ 
          {\rm delayed\, OC:} & \dtg < \dto \simeq \Dto  \end{array} \right.
\label{expect}
\end{equation}
 That the POCs shown in Figure 1 display the above features derived from $\tpk \simeq \dto$ indicates that their electron 
cooling is not dominated by iC scatterings at the T-KN transition ($n < 1$).

\begin{figure*}[t]
\centerline{\includegraphics[width=10cm,height=8cm]{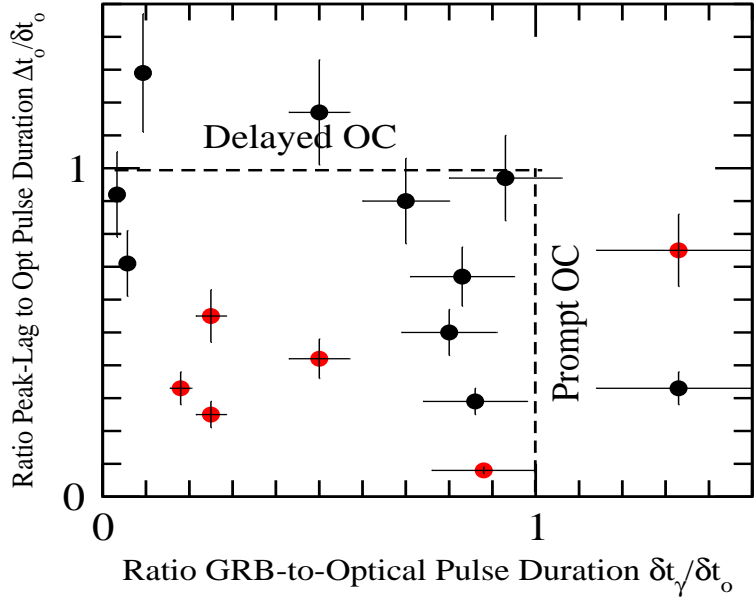}}
\figcaption{\normalsize Comparison between the temporal properties (GRB-to-OC pulse-peak lag-time $\Dto$, optical pulse 
  duration $\dto$, GRB pulse duration $\dtg$) of the GRBs and POCs of {\sl Table} \ref{data} and the {\sl expectations 
  for the POCs arising from cooling of the GRB electrons} (Equation \ref{expect}), shown with dashed lines.
  Given that two of these quantities should be equal, the diversity of POCs can be captured by plotting duration ratios 
  with same denominator for both axes. Then, we expect prompt OCs to be spread around a vertical segment of unitary length 
  at $\dtg/\dto = 1$ on the abscissa and delayed OCs to cluster around a horizontal segment of length unity at $\Dto/\dto = 1$ 
  on the ordinate.
  Black symbols are for POCs whose optical-to-GRB effective spectral slope $\bog$ is consistent with the GRB low-energy slope 
  $\bLE$ (meaning that $\bLE \leq \bog$) if POCs were produced by the cooling of GRB electrons in a constant magnetic field. 
  The error bars correspond to a plausible uncertainty $\sigma$ of 10\% for our (eye-balling) estimation of durations.
  Half of all such POCs are consistent with the temporal expectations (dashed lines) arising from the cooling of GRB electrons. 
  Within $2\sigma$, nearly all such POCs are consistent with those expectations.
  Red symbols are for POCs with peak brightness (relative to the GRB's) exceeding the expectations for electrons cooling 
  in a constant magnetic field, i.e. for OFs with $\bLE > \bog$. 
  Two-thirds of OFs are not consistent (farther than $2\sigma$) with the expectations for the GRB electron cooling
  which, together with their excessive brightness, suggests that OFs arise from a different mechanism. 
}
\end{figure*}

 For the brightest OCs, robotic telescopes may measure the OC variability associated with the GRB pulse variability.
If the latter arises from (large-scale) fluctuations in the magnetic field, then the GRB and optical flux should fluctuate
synchronously. However, such fluctuations should be easier to evidence in the OC only around its peak time, thus there
will be a bias in detecting temporally-correlated GRB and OC fluctuations mostly in prompt OCs.

 {\bf Figure 2} illustrates the diversity of POCs (from a truly prompt to a delayed OC) that is obtained by adjusting 
temporal factors (here, the duration $\ti$ of electron injection) that determine the GRB pulse duration. 
 Figure 2 also shows that GRB fluctuations (here, resulting from a variable electron injection rate $R_i$) will "survive" 
electron cooling (in the sense that they will be displayed by a variable POC light-curve) if the injection rate variability timescale
(which sets the GRB variability timescale $\dtg$) is longer than the GRB-to-OC transit-time $\tgo$ (which sets the GRB-to-optical
pulse peak lag $\Dto$ for a full electron cooling), otherwise the cooled electrons of consecutive injection episodes reach 
optically-emitting energies in short succesion, separated by a time-interval $\dtg < \tgo$, and integration over the curved 
emitting surface and over the synchrotron function will wipe-out any OC variability of timescale $\dtg$. 

 Thus, electron cooling allows the "propagation" of the GRB variability to the OC only if $\Dto < \dtg$, i.e. only for POCs.

\begin{figure*}
\centerline{\includegraphics[width=16cm,height=12cm]{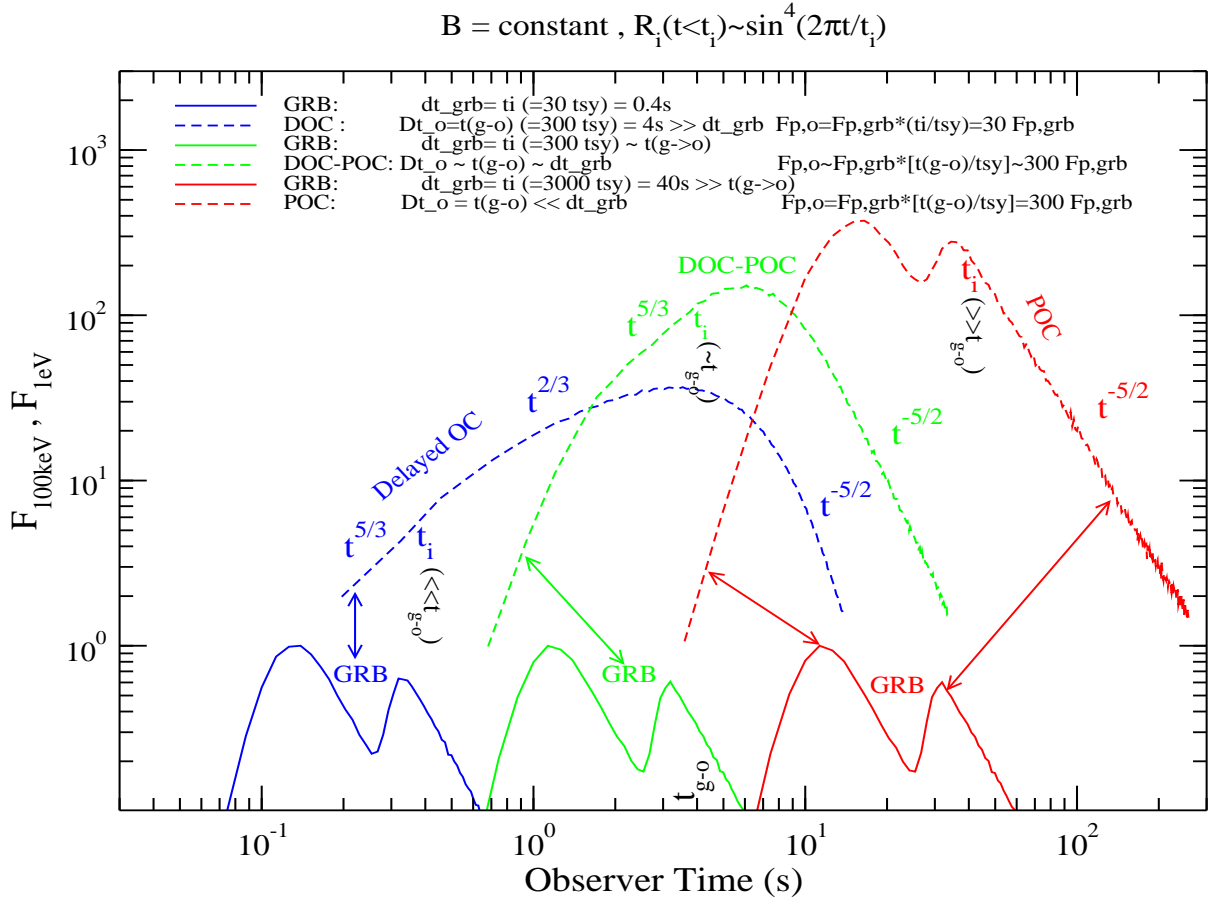}}
\figcaption{\normalsize Metamorphosis of a delayed optical counterpart (DOC) into a (truly) prompt optical counterpart (POC) 
 obtained by increasing the duration $\ti$ of electron injection (which sets the duration $\dtg$ of the GRB pulse), from well-below 
 to well-above the GRB-to-optical transit time $\tgo = 300\, \tsyi$ (which sets the GRB-to-optical peak time-delay $\Dto$). 
 Parameters are: magnetic field $B=100$ G, electrons are injected above energy $\gi = 3\ttimes 10^4$, with a $p=3$ power-law 
 distribution with energy, source Lorentz factor $\Gamma=100$. The peak energy of the $\eps \Fe$ spectrum is $\Ep \simeq 200$ keV, 
 the observer-frame SY cooling-time is $\tsyi^{(obs)} = \tsyi/(2\Gamma) = 13$ ms and the GRB-to-optical transit-time $\tgo = 4$s. 
 The GRB peak flux normalized at unity (but 1 mJy is a typical/average peak flux). The electron injection has with two sinusoidal 
 pulses and the electron cooling is SY dominated.
 For $\ti \gg \tsyi$, the GRB pulse has two peaks at $\ti/4$ and $3\ti/4$ (slightly delayed by the angular integration) and 
 the optical pulse peaks are delayed by $\tgo$.
 The legend quantifies the optical-to-GRB peak flux-ratio and the relationship between $\dtg$ and $\Dto$, with GRB and optical 
 pulses of same $\ti$ being shown with the same color. 
 That $\dtg = \ti$ and $\Dto = \tgo$ imply that: 
 $i)$ a DOC (blue), defined by $\dtg < \Dto$, is obtained for $\ti \ll \tgo$ 
     (peak-flux ratio on middle line of eq (24) in PV22,
 $ii)$ an intermediate DOC-POC (green), defined by $\dtg \simeq \Dto$, results for $\ti \simeq \tgo$, and
 $iii)$ a POC (red), defined by $\Dto \ll \dtg$, occurs for $\ti \gg \tgo$,
   with the peak-flux ratio reaching maximal value (last case in eq (24) of PV22).
 The variability timescale $\ti/2$ of the sinusoidal electron injection rate $R_i$ that is displayed by the GRB flux appears
 also in the optical if the GRB-to-optical transit time $\tgo$ is shorter than the injection variability timescale $\ti/2$,
 which means that the GRB variability is preserved for POCs (because $\Dto \sim \tgo$, $\ti \sim \dtg$ and $\tgo < \ti/2$
 imply $\Dto < \dtg$).
 For $\ti \ll \tgo$, the two sines of injected electrons cool and reach optical-emitting energies separated by $\ti/2 \ll \tgo \simeq \Dto$,
 with integration over the synchrotron function and the angular opening of the ejecta ironing out the initial variability,
 which means that GRB variability is lost for DOCs (similar to above, $\ti < \tgo$ implies $\dtg < \Dto$).
 Thus, {\sl delayed OCs lose the GRB variability, but prompt OCs retain it}.
 The indicated OC flux power-laws are those of eq (21) in PV22 : $f_\eps (t < \ti/2) \sim t^{5/3}$, 
 $f_\eps (\ti/2 < t < \tgo) \sim t^{2/3}$ with $\ti/2$ marking the end of the first injection episode, 
 and are close to those displayed by the rise of the OC flux calculated numerically despite that the first analytical result
 (for $t < \ti/2$) was derived for a constant injection rate. 
 (The $f_\eps \sim t^4$ earliest rise shows the injection rate $R_i$, the emission from the cooling tail being overshined by 
   the sharp rise of $R_i$).
 The OC power-law falling flux is the LAE given in eq (26) of PV22 for the slope $\beta = -1/2$ of the integrated spectrum 
 of the SY emission from a quasi-monoenergetic cooled electron distribution.
}
\end{figure*}

\vspace{2mm}
\subsection{\bf Spectral Correlations between GRB and POC}
\label{spekcorrel}

 For a constant magnetic field, the electron cooling (through any process) yields an effective optical-to-GRB slope
harder than the GRB low-energy slope, $\bog \geq \bLE$, with the equality resulting if either electrons do not cool 
significantly during the magnetic field life-time $\tB$, leading to $\bLE = \bog = \beta_o = 1/3$ with $\beta_o$ 
being the optical continuum slope, or if electrons cool below optical during $\min(\ti,\tB)$, leading to 
$\bLE = \bog = \beta_o = -(n-1)/2$. If either the magnetic field life-time or electron injection duration satisfy 
$\tci < \tB,\ti < \tgo$, then the SY emission integrated spectrum peaks between optical and $\gamma$-rays and 
$\bog > \bLE -(n-1)/2$ ($\beta_o = 1/3$ or $\beta_o = -(n-1)/2$).

 For a minimal electron cooling ($\tB < \tci$, $\bLE = 1/3$), one expects 
$i)$ a very POC (case 1 in {\sl Tables} \ref{ALL} and \ref{TKN}) because the magnetic field life-time 
  sets the pulse duration and peak-epoch at any energy, and 
$ii)$ an POC dimmer than the GRB by a factor $(\Ep/1 eV)^{1/3} \simeq 40$ or about 4 magnitudes. 
 For a typical GRB average flux of 1 mJy, the POC should be of magnitude $R=20$; for a bright GRB with an average flux
of 10 mJy, one gets $R=18$. Even for the latter case, the POC, occuring during the burst, is too dim to be
monitored by robotic telescopes at such early times, thus there will be a {\sl bias} against following-up POCs arising 
from the cooling of electrons in GRBs with a hard low-energy slope. When such POCs are detected (as for GRB 990123 
-- {\sl Table} \ref{data}), it is more likely that that optical emission arises from another mechanism, a possibility
that can be tested (\S\ref{identify}).

 If electrons cool well below gamma-rays ($\tB > \tci$, $\bLE = -1/2$ for SY-dominated electron cooling), the POC may 
be either prompt or delayed, and the POC may be brighter than the GRB by a factor up to $(\Ep/1 eV)^{1/2} \simeq 300$, 
or about 6 magnitudes. For an average GRB flux of 1 mJy, the maximal POC brightness is $R=10$, while brighter bursts, 
with an average flux of 10 mJy, could yield an POC of $R=7.5$, thus the cooling of GRB electrons may account for the 
OFs of bright bursts such as GRB 090123, 080319B, or 130427A ({\sl Table} \ref{data}).

 The above considerations suggest a {\sl correlation} between: 

$i)$ {\sl POC type and GRB low-energy slope} $\bLE$, induced by GRBs with a short magnetic field life-time $\tB < \tci$: 
 harder slopes $\bLE$ should be associated more often with POCs than with delayed OCs. All POCs in {\sl Table} \ref{data} 
 have soft low-energy slopes $\bLE \in (-0.5,-0.3)$ in a narrow range, thus this correlation cannot be tested with the POCs 
 identified here. 

$ii)$ {\sl optical-to-GRB $\bog$ and GRB low-energy $\bLE$ slopes}, induced by their dependence on the magnetic field 
 life-time $\tB$, which can be quantified using equations (30), (34), (48)--(50) of PV22 for the slope $\bLE$ and 
 Equations (\ref{bogsy})--(\ref{bogic}) for the effective slope $\bog$.  
 For SY-dominated electron cooling, the slope $\bog$ on the 2nd and 4th branches of Equation (\ref{bogsy}) show that a 
 magnetic field life-time $\tB \in (0,10)\,\tsyi$ yields correlated slopes $\bog \in (0,1/3)$ and $\bLE \in (-1/2,1/3)$,
 while for $\tB > 10\,\tsyi$, the resulting range of slopes $\bog \in (-1/2,0)$ is uncorrelated with the only slope
 $\bLE = -1/2$.

 In general, this correlation can be written as a condition: $\bLE \geq \bog$, meaning that {\sl GRBs with harder low-energy 
 slopes are associated more often with dimmer POCs}. If the POC emission from cooling GRB electrons is not overshined by another
 mechanism, then the observational bias against following such dimmer POCs will lead to a paucity of GRBs with hard 
 low-energy slopes and POCs, which accounts for softness of the GRBs with the POCs listed in {\sl Table} \ref{data}.  

$iii)$ {\sl POC type and POC-to-GRB relative brightness} (quantified by the effective slope $\bog$), with POCs 
 being dimmer on average than delayed ones. This correlation is supported by the POCs of {\sl Table} \ref{data}, with 
 the average slope $\bog$ of the former being harder than for the latter. 

{\sl Optical extinction and reddening by dust in the host galaxy}.
 Correlations involving the POC brightness may be weakened by the hard-to-determine accurately dust-extinction in the 
host galaxy, with an extinction $A_{V,h}$ in the host galaxy frame reducing the POC brightness by about $(1+z)A_{V,h}$ 
magnitudes (for a linear reddening curve).  
 Thus, dust-extinction hardens the slope $\bog$ by $\delta \bog = 0.4(z+1)A_{V,h}/5$. 
 For $A_{V,h} = 1$ mag and a redshift $z=2$, the resulting hardening $\delta \bog = 1/4$ is about one-third of the 
entire expected range $\bog \in (-1/2,1/3)$.

 Furthermore, dust extinction is accompanied by a softening of the optical continuum slope $\beta_o$ by $\delta \beta_o = 
-0.4(z+1) A_{V,h}$, i.e. $\delta \beta_o = -1.2$ for $z=2$ and $A_{V,h} = 1$ mag, thus a hard intrinsic optical slope 
$\beta^{(intr)}_o = 1/3$ (expected for a magnetic field that lives shorter than the GRB-to-POC transit-time, $\tB < \tgo$) 
could become a softer measured slope $\beta^{(dust)}_o \simeq -1$.

\vspace{2mm}
\subsection{\bf Identification of POCs from Cooling of GRB Electrons}
\label{identify}

 Thus, the {\sl temporal} correlation between POC type and the optical-to-GRB pulse duration ratio $\dto/\dtg$ and 
the {\sl spectral} condition $\bog > \bLE$ between the optical-to-GRB effective slope and the GRB low-energy slope 
are two criteria to identify the POCs originating from the cooling of GRB electrons. Because the spectral criterion
is only an inequality, that criteerion is rather weak and could yield many "false positives", as POC arising from
other mechanisms may satisfy it.

 {\bf Figure 1} and {\bf Table \ref{data}} show that {\sl most OFs}, defined by $\bog < \bLE$ (i.e. POCs that 
are brighter than expected from the cooling of GRB electrons in a constant magnetic field), {\sl do not satisfy the temporal 
correlation expected for POCs from electron cooling} ($\dto \simeq \dtg$ for POCs, $\Dto \simeq \dto$ for delayed OCs). 

 We note that an increasing magnetic field could account for the higher brightness of OFs even when they 
arise from the cooling of GRB electrons, and that such an increasing magnetic field should not invalidate the expected 
temporal correlations for each POC type because those correlations arise from that the electron cooling-time at energy 
$\eps$ is comparable to the electron transit-time to energy $\eps$, which is correct even for a variable $B$. 
 This implies that the temporal criterion for the identification of POCs from the cooling of GRB electrons works even
for a variable magnetic field, while the spectral criterion is useful only for a constant magnetic and may miss true POCs
from GRB electron cooling if the magnetic field is increasing. 

 Thus, the temporal criterion is clearly superior to the spectral one in selecting POCs that arise from the cooling of 
GRB electrons, but that does not mean that the spectral criterion is useless because the temporal criterion has a 
limitation that is alleviated by adding the spectral criterion.
 Because the temporal correlations induced by electron cooling arise solely from the pulse peak-epoch being comparable 
to the pulse duration ($\tpe \simeq \dteps$), it is not sensitive to the origin of the cooling electrons and cannot 
discriminate among various mechanisms that produce optical emission (pairs produced by GeV photons, reverse-shock, 
SY emission in synchrotron self-Compton GRBs) {\sl if} the timing of POC pulse is set by cooling of electrons.
 
 If the POC arises from another mechanism overshining the emission from cooling GRB electrons, then the POC will not
satisfy the spectral condition $\bog > \bLE$. 
 Consequently, adding the spectral condition to the temporal criterion is a trade-off, as it increases the probability 
that an POC staisfying both criteria arises from the cooling of GRB electrons, at the risk of missing some true POCs 
from GRB electrons cooling in an increasing magnetic field.

 Putting together the considerations in \S\ref{tempcorrel}, one can conclude that {\sl GRB variablity is passed on the 
OC only for POCs}, with that being either an observational bias (for GRB variability arising from fluctuations in the 
magnetic field) or a consequence of electron cooling (for GRB variability from fluctuations in the injection rate). 
To the extent that OC variability can be measured by robotic telescopes, this conclusion provides another criterion 
for identifying OCs arising from the cooling of GRB electrons.

\vspace{3mm}
\section{\bf Conclusions}
   
 The aim of this work is to investigate what new information can be extracted from the properties of the
Prompt Optical Counterparts resulting from cooling of GRB electrons. 

 Starting from the durations $\dtg$ and $\dto$ of GRB and POC pulses (which are set by the corresponding electron cooling 
times for the emission from a bright-spot, the angular time-spread $\tang$ for the emission from a uniform-brightness surface 
or for AD-cooling, or the two temporal parameters $\tB$ and $\ti$ for the duration of magnetic field and electron injection)
and the lag-time $\Dto$ between the peaks of the GRB and POC pulses (which is set by the gamma-to-optical transit-time $\tgo$ 
or by the timescale $\tB$), it can be shown that the POCs arising from the cooling of GRB electrons are of two types:
"prompt", for which $\Dto < \dtg = \dto$, and "delayed", for which $\dtg < \Dto = \dto$. 

 The preceeding inequalities are definitions and the following equalities represent a test for the electron-cooling 
model for POCs. Adding the condition $\bog \geq \bLE$ between the POC-to-GRB effective spectral slope $\bog$ and the 
GRB low-energy slope $\bLE$ strengthens this temporal criterion for identifying POCs arising from the cooling of GRB 
electrons by discriminating the POCs arising from other mechanisms (involving electron cooling or not), although that 
may miss some POCs from cooling if GRB electrons if the magnetic field were to increase. 

 Adding that POCs should be associated more often with GRBs with a harder low-energy slope $\bLE$, these two
correlations imply that POCs should be dimmer (on average) than delayed ones. This correlation finds support
in a set of ten POCs, with the average POC-to-GRB brightness ratio $\bog$ being harder for POCs.

 If GRB electrons that do not cool significantly during the magnetic field life then the burst should have a hard 
low-energy slope $\bLE=1/3$ and the POC should be prompt and dimmer be a factor $(100keV/1keV)^{1/3}$ (about 4 magnitudes) 
than the burst. Thus, a burst of average flux of 1 (10) mJy will be accompanied by an POC of magnitude $R=20$ (18),
which implies that some GRBs with a hard low-energy slope $\bLE$ will be accompanied by a prompt and dim intrinsic POC 
emission (from cooling of GRB electrons) that is dificult to detect with robotic telescopes. That sets a bias against 
detecting the intrinsic POCs of hard GRBs.
  
 If GRB electrons cool to optical energies then the burst should have a soft low-energy slope $\bLE = -1/2$ 
(for SY-dominated electron cooling) and the POC may be brighter than the burst by a factor up to $(100keV/1keV)^{1/2}$ 
(6 magnitudes). Thus, the POC of a burst with a soft slope $\bLE$ and average flux of 1 (10) mJy could be as bright as 
magnitude $R=10$ (7.5), which is comparable to the brightness of the OFs accompanying GRBs 990123, 080319B, 
and 130427A ({\sl Table} \ref{data}). However, all those bursts have a hard low-energy slope $\bLE > 0$, which 
implies an incomplete electron cooling and yields a dim intrinsic POC emmission (from the cooled GRB electrons), 
thus their OFs must have arose from another mechanism (emission from pairs formed from the GeV prompt 
emission, external reverse-shock, SY emission in the synchrotron self-Compton model for GRBs).

\end{document}